\documentclass[%
notitlepage,%
onecolumn,%
oneside,%
floats,%
aps,%
prd,%
nobibnotes,%
nofootinbib,%
amsmath,%
amssymb,%
amsfonts,%
amscd,%
superscriptaddress,%
eqsecnum%
]{revtex4-1}

\usepackage[utf8]{inputenc}
\usepackage{graphicx, array, dcolumn}
\usepackage[paperwidth=210mm, paperheight=297mm, centering, hmargin=2.0cm, vmargin=2.4cm]{geometry}
\usepackage{ulem}

\begin{document}
\title{Relic abundance of MeV millicharged particles}

\author{A.D.\,Dolgov}
\email{dolgov@fe.infn.it}
\affiliation{Novosibirsk State University, Novosibirsk, 630090, Russia}
\affiliation{Institute for Theoretical and Experimental Physics, Moscow, 117218, Russia}
\affiliation{Dipartimento di Fisica, Universit\`a degli Studi di Ferrara, I-44100 Ferrara, Italy}

\author{A.S.\,Rudenko}
\email{a.s.rudenko@inp.nsk.su}
\affiliation{Novosibirsk State University, Novosibirsk, 630090, Russia}
\affiliation{Budker Institute of Nuclear Physics, Novosibirsk, 630090, Russia}

\begin{abstract}

The relic abundance of light millicharged particles (MCPs)
with the electric charge $e' = 5\cdot 10^{-5} e$ and with the mass slightly below or above the electron
mass is calculated. The abundance depends on the mass ratio $\eta=m_X/m_e$ and for $\eta<1$
can be high enough to allow MCPs to be the cosmological dark matter or to
make a noticeable contribution to it. On the other hand, for $\eta \gtrsim 1$
the cosmological energy density of MCPs can be quite low,
$\Omega_X h_0^2 \approx 0.02$ for scalar MCPs, and $\Omega_X h_0^2 \approx 0.001$ for spin 1/2 fermions.
But even the lowest value of $\Omega_X h_0^2$ is in tension
with several existing limits on the MCP abundances and parameters.
However, these limits have been derived under some natural or reasonable assumptions on
the properties of MCPs. If these assumptions are relaxed,
a patch in the mass-charge plot of MCPs may appear, permitting them to be dark matter particles.

\end{abstract}

\maketitle

\section{Introduction}
Millicharged particles (which we will denote either as MCP or $X$) are hypothetical particles with
an electric charge $e'=\epsilon e$, which is much smaller than the elementary
charge $e$, i.e. $\epsilon \ll 1$. The possibility of existence of such particles was
suggested many years ago by different authors, e.g. in~\cite{ignatiev} in
connection with a possible nonconservation of the electric charge, and in~\cite{holdom} 
in the model with a second $U(1)$ gauge field, the "paraphoton".
Later in~\cite{goldberg} particles with $\epsilon \ll 1$ were considered as
candidates for dark matter. Since then millicharged particles were widely discussed
in the literature, and various constraints on their parameters (mass and charge)
were obtained from laboratory and accelerator experiments as well as from
astrophysical and cosmological considerations. The plots with the excluded regions
of mass-charge parameter space one can find e.g. in~\cite{dobroliubov,
davidson, davidson-2, SLAC, davidson-3, dubovsky, positronium}.
The recent limits allow only for a minor cosmological fraction of MCPs, the best one
up to now is $\Omega_X h_0^2 <  0.001$ (95\% CL)~\cite{dubovsky-2}.

However, all the limits are derived under some minimal assumptions on the MCP properties.
The only thing which is taken for sure is the MCP electromagnetic interaction with a tiny
electric charge.  By default all (or almost all) other possible interactions of MCPs are neglected.
We will not use these assumptions and take instead the maximum freedom principle, i.e. assume that
anything which is not explicitly excluded is permitted.
If this is the case, then MCPs can make up the whole cosmological dark matter or be a 
considerable~fraction~of~it.

Our paper is organized as follows. In Sect.~\ref{sect-mot} we explain in more detail
motivation of the study. In Sect.~\ref{sect-boltzmann} we discuss the kinetic
equation which describes the number of $X$-particles in expanding universe and
analytically calculate their cosmological abundance for the case of $m_X$ slightly above
and slightly below $m_e  = 0.511 $ MeV
(subthreshold annihilation). In Sect.~\ref{sect-num} we present the results of the numerical computations.
Finally, in Sect.~\ref{sect-conclud} we conclude.

\section{Motivation} \label{sect-mot}

The idea that millicharged particles constitute a part of dark matter is quite
intriguing. For instance, it was proposed recently~\cite{bdt} that the mystery of
the origin of galactic magnetic fields could be solved if one assumed that these
fields were created due to interaction between electrons and millicharged dark
matter particles.

The particles much lighter than 1 MeV and with the charge,
$\epsilon \gtrsim 10^{-9}$, are supposed to be excluded by their impact on Big Bang
Nucleosynthesis (BBN)~\cite{davidson-3}. However, an agreement with observations can
be restored if the cosmological lepton asymmetry is non-vanishing~\cite{bdt-2}. Nevertheless,
we will not consider here very light millicharged particles with $m_X \ll 1$ MeV.

The limits on $\epsilon$ are obtained also from invisible decays of
orthopositronium. Such limits are applied only for $m_X<m_e$ and constitute $\epsilon < 3.4 \cdot 10^{-5}$ for $m_X<<m_e$
\cite{positronium}. However, they are quite weak for $m_X \lesssim m_e$, because
the decay probability is proportional to
$(1-m_X^2/m_e^2)^{n/2}$ (here $n=1$ for spin 1/2 $X$-particles and $n=3$ for spin 0)~\cite{gninenko},
and  for $m_X \geq m_e$  no constraint on $\epsilon$ can be derived at all.

The experiment uniquely suited to the detection of millicharged particles was
performed at SLAC~\cite{SLAC}, where the following bounds (95\% CL) were obtained:
$\epsilon < 2.0\cdot 10^{-5}$ for $m_X=0.1$ MeV, and $\epsilon < 4.1\cdot
10^{-5}$ for $m_X=1$ MeV. However, without the assumption of linearity of the
scintillator for very small energy depositions, the bounds are a factor of 2 less
stringent~\cite{SLAC}. Roughly speaking, according to this experiment the electric charge of MCPs is
bounded from above by $\epsilon \lesssim 5\cdot 10^{-5}$ for $m_X \sim m_e$.
The SLAC bounds are valid if MCPs have sufficiently weak interaction with the usual matter
to propagate 110 meters of the sandstone between the source and detector. This is true if
the only interaction of MCPs is the milli-electromagnetic one. New stronger interactions could
destroy the bound. Some related references can be found in the list~\cite{anom-MCP},
where restrictive bounds are derived but probably more exotic options still remain open.

There is also the region $10^{-9} \lesssim \epsilon \lesssim 10^{-7}$, $m_X
\lesssim 5$ MeV which is excluded by consideration of the energy-loss rate of the Supernova
1987A~\cite{davidson-3}, but we are interested here in larger allowed values of $\epsilon$.

Besides the discussed results, the millicharged particles with $\epsilon \lesssim
5\cdot 10^{-5}$ and $m_X \sim m_e$ are reported to be excluded by the  limits on
their relic abundance, $\Omega_X h_0^2$. The corresponding exclusion plot can be
found e.g. in~\cite{dubovsky}. Therein the limits were obtained using the
Lee-Weinberg formula \cite{lee-weinberg} for the relic abundances, precise cosmic
microwave background (CMB) data from WMAP, and  the standard BBN scenario.
However, details of calculation, namely how the bound on $\Omega_X h_0^2$ was
translated into the constraints on mass and charge of $X$-particles, were not given
there. Therefore, we believe that it is necessary to revisit these bounds with more
accuracy. So, in this paper we calculate carefully
the relic abundance of millicharged particles with masses not very different from the
electron mass, $m_X \sim m_e$. We consider not only the usual case when
$X$-particles can annihilate into lighter particles, but also the annihilation
in $e^+e^-$ "below threshold" (if $m_X < m_e$) which is allowed for energetic
$X$-particles from the tail of their energy distribution \cite{gondolo, griest, agnolo}.
To the best of our knowledge, subthreshold annihilation of MCPs was not studied before.

As it has been mentioned above, the most stringent bound is obtained from the analysis of the
angular fluctuation spectrum of CMB performed in~\cite{dubovsky-2}. An essential point in the
derivation of this bound is the assumption that MCPs and protons with electrons are strongly coupled
to each other, so they oscillate as a unique substance creating the acoustic oscillations of the photon
temperature. So the shape of the angular spectrum is determined by the sum of the cosmological densities
of protons and MCPs. Separate measurement of the diffusion (Silk) damping of high multipoles allows to
separate the contribution of protons and MCPs and to obtain the record bound on the cosmological
abundance of MCPs. The assumption of a sufficiently strong coupling between protons and MCPs
is based on the estimate of the Coulomb interactions  between them. This estimate is valid if the temperatures
of the protons and MCPs are the same and equal to the temperature of the CMB photons.  However, one
can imagine a scenario, e.g. with some new particles interacting with MCPs, when the MCP temperature
could be noticeably higher than $T_\gamma$. In this case the Coulomb coupling drops down, MCPs
do not participate in the proton acoustic oscillations, and the bound is destroyed.
This scenario will be studied elsewhere.

\section{Boltzmann equation} \label{sect-boltzmann}

When the universe was hot enough, millicharged particles were in thermal
equilibrium, if $\epsilon$ was not vanishingly tiny.
For example, for the temperatures larger than $m_X$,
equilibrium with respect to the elastic MCP scattering off electrons,
as well as $X \bar X$ annihilation into $e^+ e^-$ pairs,
was established at $T \lesssim \alpha^2 \epsilon^2 m_{Pl} \sim 10^5 (\epsilon /10^{-5})^2 $ GeV.
However, the universe was expanding and after the
moment when the annihilation rate of $X\bar{X}$ pairs
became smaller than Hubble parameter, $\Gamma_{ann}
\lesssim H$,  due to the Boltzmann suppression of
their number density  at $T<m_X$,
the $X$-particle annihilation practically stopped and their number became
constant in comoving volume. This phenomenon is called {\it freeze-out}.

The derivation of Boltzmann (kinetic) equation which describes the evolution of
particle number density before and after decoupling is discussed in detail in~\cite{kolb, gondolo}. 
Here we briefly remind this derivation and assumptions under which it is valid.

Let us assume there are no other particles beyond the Standard Model (SM) ones,
except MCPs, which can be scalars or fermions. Such $X$-particles
are supposed to be stable and interacting directly only with photons and via photons
with other charged particles of SM. It is also implicitly assumed that $\epsilon$, though small, but
still is large enough, so the specified below conditions are fulfilled.

\begin{figure}[h]
\center
\includegraphics[scale=0.75]{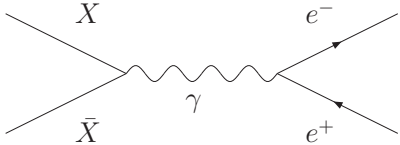}
\caption{\label{fig:1} The Feynman diagram for $X\bar{X} \to e^+e^-$ annihilation}
\end{figure}

The temperature of freeze-out, $T_f$, should be much smaller than $m_X$,
otherwise MCPs would overclose the universe.
For $m_X \sim m_e$ the production and annihilation of MCPs proceeded mainly through the reaction
$X\bar{X} \leftrightarrow e^+e^-$, the cross-section of which at low energies is
$\sim \epsilon^2 \alpha^2/ m_X^2$ (see Fig.~\ref{fig:1}).
Other inelastic processes involving $X$-particles
are inessential at such temperatures, e.g. the cross-section of two-photon
annihilation $X\bar{X} \rightarrow \gamma\gamma$ is heavily suppressed as
$\epsilon^4 \alpha^2$, and the plasmon decay $\gamma_P \to X\bar{X}$ is also
ineffective for $m_X \sim m_e$ (plasmon decay is operative at $T\gtrsim 10$ MeV
since the plasmon ''mass'' (plasma frequency) $\omega_P \sim 0.1 T$ must be $\gtrsim 2m_X$)~\cite{ber-lep}.
One can consider also the three-particle
annihilation $X\bar{X}e^\pm \rightarrow e^\pm \gamma$, the cross-section of which is
suppressed only as $\epsilon^2 \alpha^3$, but it has an additional phase space suppression
with respect to the two-body channel $X\bar{X} \to e^+e^-$.
However, such three-particle reaction may be essential for temperatures and
masses of MCPs lower than ones considered here,
when ordinary annihilation $X\bar{X} \to e^+e^-$ is heavily suppressed.

It is usually assumed that kinetic equilibrium of $X$-particles was maintained even
after freezing of their annihilation, i.e. at temperatures $T \lesssim m_X$. The equilibrium
remained while the rate of elastic scattering $X+e^- \to X+e^-$ was much larger than the rate of
the universe expansion, $H$. If the asymmetry between particles and antiparticles was not
too high, the particle occupation numbers were small, $f \ll 1$, and thus at this period
$X$-particles, electrons, and positrons obeyed the Boltzmann statistics and were
described by the equilibrium distribution functions, $f_i^{eq}={\rm exp}[-(E_i-\mu_i)/T]$,
with $E_i$  and $\mu_i$  being respectively the energy and the chemical potential of
$i$-th sort of particles.

Chemical potentials are introduced to describe a difference between the number densities of
particles and antiparticles in thermal equilibrium. The evolution of $\mu_j$ is governed by inelastic reactions and in full
equilibrium they satisfy the condition $\mu + \bar \mu = 0$, where $\bar \mu$ is the chemical
potential of antiparticles. If the chemical equilibrium is not maintained, the above condition is not
necessarily fulfilled and chemical potentials may be non-zero even in absence of particle-antiparticle
asymmetry. Usually it is assumed that the number density of $X$-particles is equal
to that of antiparticles, $ n_X = n_{\bar X} $. When the annihilation of $ X \bar X $ was frozen but kinetic equilibrium
is maintained, the particle and antiparticle distributions have the same form $f = \exp [(\mu-E)/T]$ with
$\mu = \bar \mu$. In this case $\mu$ is often called {\it effective chemical potential}, which is usually a function of time but
not of the particle energy. Correspondingly the distribution can be presented as
\begin{equation}
f = C(t) f^{(eq)} \equiv  C(t) \exp (- E/T) .
\label{f-mu}
\end{equation}

We also assume that $T$-invariance holds, therefore amplitudes squared and summed
over spins are equal for direct and reverse reactions, $\sum_s |A_{a \to b}|^2=
\sum_s |A_{b \to a}|^2$, with an evident change of the signs of velocities of the participating particles.

Finally, under all these assumptions the Boltzmann equation for $X$-particles in
FLRW metric can be written as
\begin{equation} \label{Boltzmann}
\dot{n}_X+3Hn_X=-<\sigma v>(n_X^2-n^2_{X, eq}).
\end{equation}
Here $n_X$ is the number density of $X$-particles, $n_{X, eq}$ is the equilibrium
one, i.e. having vanishing chemical potential,
$H=\dot{a}/a$ is the Hubble parameter ($a$ is the cosmological scale factor),
and $<\sigma v>$ is thermally averaged cross-section of process
$X\bar{X}\to e^+e^-$ times the M\o ller velocity.

Following~\cite{gondolo} one can obtain that
\begin{equation} \label{cross-sect-aver}
<\sigma v>=\frac{1}{8m_X^4TK_2^2(m_X/T)}\int_{max(4m_X^2, 4m_e^2)}^\infty
\sqrt{s}(s-4m_X^2)\sigma(s)K_1\left(\frac{\sqrt{s}}{T}\right)ds,
\end{equation}
where the cross-section $\sigma(X\bar{X}\to e^+e^-)$ is summed over final and
averaged over initial spins, $K_i$ are the modified Bessel functions of second kind
and order $i$. Note that the cases $m_X \leq m_e$ and $m_X \geq m_e$ differ from
each other only in lower limit of integral (\ref{cross-sect-aver}).

Instead of $n_X(t)$ let us introduce the dimensionless quantity $Y(\xi)=n_X/s$. Here
$\xi=m_X/T$ and $s=g_{*s}(2\pi^2/45)T^3$ is the entropy density (it should not be
confused with the Mandelstam variable $s$), where $g_{*s}=3.94$ is the effective number
of relativistic species in the entropy density for $T \ll m_e$,
which includes photons and three species of massless neutrinos.
The function $Y(\xi)$ is
very convenient because of the relation $\dot{s}+3Hs=0$, which means that total entropy
in comoving volume is conserved, $sa^3=const$.
Moreover, $Y(\xi)$ has a simple
physical meaning, it is proportional to the total number of $X$-particles,
$Y=n_X/s\sim n_X a^3=N_X$.

In radiation-dominated universe ($T \gtrsim 1$ eV) the energy density is
\begin{equation} \label{rho}
\rho=\frac{3}{32\pi}\frac{m_{Pl}^2}{t^2}=g_* \frac{\pi^2}{30}T^4,
\end{equation}
where $m_{Pl}\approx 1.22 \cdot 10^{19}$ GeV is the Planck mass, $g_*=3.38$ is the
effective number of relativistic species in the cosmological energy density for $T \ll m_e$. The
relation (\ref{rho}) shows that temperature $T$ behaves like $1/\sqrt{t}$. Now it is
straightforward to change variable $t$ to $\xi=m_X/T$.

The Boltzmann equation (\ref{Boltzmann}) then becomes
\begin{equation}
\frac{dY}{d\xi}=-\sqrt{\frac{45}{4\pi^3g_*}}\frac{m_{Pl}}{m_X^2}\xi s<\sigma
v>(Y^2-Y^2_{eq}),
\end{equation}
where $Y_{eq}=n_{X, eq}/s$, and $n_{X,
eq}=(g_X/2\pi^2)K_2(m_X/T)m_X^2T$~\cite{gondolo} (here $g_X$ is the number of spin
degrees of freedom for $X$-particles, $g_X=1$ for scalars and $g_X=2$ for spin 1/2 fermions).

\vspace{3mm}

I. When $X$-particles are scalars, the cross-section of $X\bar{X}\to e^+e^-$
annihilation for non-identical $X$ and $\bar{X}$ is equal to:
\begin{equation} \label{cross-sect-scal}
\sigma(X\bar{X}\to
e^+e^-)=\frac{4\pi}{3}\alpha^2\epsilon^2\frac{1}{s}\sqrt{1-\frac{4m_e^2}{s}}\sqrt{1-\frac{4m_X^2}{s}}\left(1+\frac{2m_e^2}{s}\right).
\end{equation}
The factor $\sqrt{1-4m_e^2/s}$ comes as usually from the phase space of the final
particles and the factor $\sqrt{1-4m_X^2/s}$ originates here from the initial c.m. velocity related to
the scalar particles annihilation in $P$-wave. Therefore, this cross-section vanishes for
$s=4m_X^2$ and/or $s=4m_e^2$.

Substituting (\ref{cross-sect-scal}) into (\ref{cross-sect-aver}) one finds
\begin{equation} \label{cross-sect-aver-scal}
<\sigma
v>=\frac{4\pi}{3}\alpha^2\epsilon^2\frac{1}{m_X^2}\frac{\xi}{\eta^3K_2^2(\xi)}I_s(\xi,\eta),
\end{equation}
where
\begin{equation}
I_s(\xi,\eta)=\int_{max(1,\eta^2)}^\infty
\sqrt{x-1}\left(1-\frac{\eta^2}{x}\right)^{3/2}\left(1+\frac{1}{2x}\right)K_1\left(\frac{2\xi\sqrt{x}}{\eta}\right)dx
\end{equation}
with $x=s/4m_e^2$, $\eta=m_X/m_e$.

\vspace{3mm}

II. When $X$-particles are spin 1/2 fermions,
\begin{equation} \label{cross-sect-ferm}
\sigma(X\bar{X}\to
e^+e^-)=\frac{4\pi}{3}\alpha^2\epsilon^2\frac{1}{s}\frac{\sqrt{1-4m_e^2/s}}{\sqrt{1-4m_X^2/s}}\left(1+\frac{2m_e^2}{s}+\frac{2m_X^2}{s}+\frac{4m_X^2m_e^2}{s^2}\right).
\end{equation}
For $m_X=m_e=m$ this cross-section is a finite constant,
$\sigma=3\pi\alpha^2\epsilon^2/4m^2$ at threshold $s=4m^2$.

Substituting (\ref{cross-sect-ferm}) into (\ref{cross-sect-aver}) one has
\begin{equation} \label{cross-sect-aver-ferm}
<\sigma
v>=\frac{4\pi}{3}\alpha^2\epsilon^2\frac{1}{m_X^2}\frac{\xi}{\eta^3K_2^2(\xi)}I_f(\xi,\eta),
\end{equation}
where
\begin{equation}
I_f(\xi,\eta)=\int_{max(1,\eta^2)}^\infty
\sqrt{x-1}\left(1-\frac{\eta^2}{x}\right)^{1/2}\left(1+\frac{1}{2x}+\frac{\eta^2}{2x}+\frac{\eta^2}{4x^2}\right)K_1\left(\frac{2\xi\sqrt{x}}{\eta}\right)dx.
\end{equation}

Finally, the Boltzmann equation takes the following form
\begin{equation} \label{Boltzmann-final}
\frac{dY}{d\xi}=-\frac{\epsilon_5^2}{\eta^4}I(\xi,\eta)\left(a\frac{Y^2}{\xi
K_2^2(\xi)}-b\xi^3\right),
\end{equation}
where $\epsilon_5=10^5\cdot\epsilon$; $I(\xi,\eta)=I_s$ when $X$ is
a scalar, and $I(\xi,\eta)=I_f$ when $X$ is a spin 1/2 fermion; $a$ and $b$ are the numerical constants:
\begin{equation}
a=10^{-10}\cdot\frac{4\sqrt{5}\pi^{3/2}\alpha^2}{45}\frac{m_{Pl}}{m_e}
\frac{g_{*s}}{\sqrt{g_*}}\approx 3.0\cdot 10^8,
\end{equation}
\begin{equation}
b=10^{-10}\cdot\frac{45\sqrt{5}\,\alpha^2}{4\pi^{13/2}}\frac{m_{Pl}}{m_e}\frac{g_X^2}{g_{*s}\sqrt{g_*}}\approx
2.6\cdot 10^5\cdot g_X^2.
\end{equation}

As it is mentioned above, due to the universe expansion the reaction $X\bar{X}\to
e^+e^-$ eventually "freezes out". After that the number of stable $X$-particles
remains constant in the comoving volume, and nowadays their energy density tends to:
\begin{equation}
\rho_{X,0}=m_X \cdot n_{X,0}=m_e\eta \cdot Y_0 s_0.
\end{equation}
Here the present-day values are marked with the index 0, $s_0 \approx 2.9\cdot 10^3$
cm$^{-3}$ is the present entropy density, and we take into account that massive
$X$-particles are nonrelativistic today.

The contribution of such $X$-particles to the cosmological energy density
constitutes
\begin{equation} \label{omega}
\Omega_X h_0^2=\frac{\rho_{X,0}}{\rho_c}h_0^2 \simeq 1.4\cdot 10^5 \eta Y_0,
\end{equation}
where $\rho_c \approx 1.88\cdot10^{-29}h_0^2$ g/cm$^3$ is the critical density of
universe.

\section{Numerical calculation} \label{sect-num}

Here we solve  the Boltzmann equation~(\ref{Boltzmann-final}) numerically in
order to calculate $Y_0$ and accordingly $\Omega_X h_0^2$. Obviously, the weaker
interaction of $X$-particles is, the larger number of them remains after their
decoupling, i.e. a smaller $\epsilon_5$ corresponds to a larger $\Omega_X
h_0^2$. We use throughout the calculation the value $\epsilon_5=5$, which
corresponds the upper bound on $\epsilon$ obtained at SLAC~\cite{SLAC}, though
as we mentioned above this limit may be questioned if some anomalous interactions of $X$
are effective. Since the value of the electric charge of $X$-particles is fixed now,
there remains only one free parameter, $\eta=m_X/m_e$.

\begin{figure}[h]
\center
\includegraphics[scale=1]{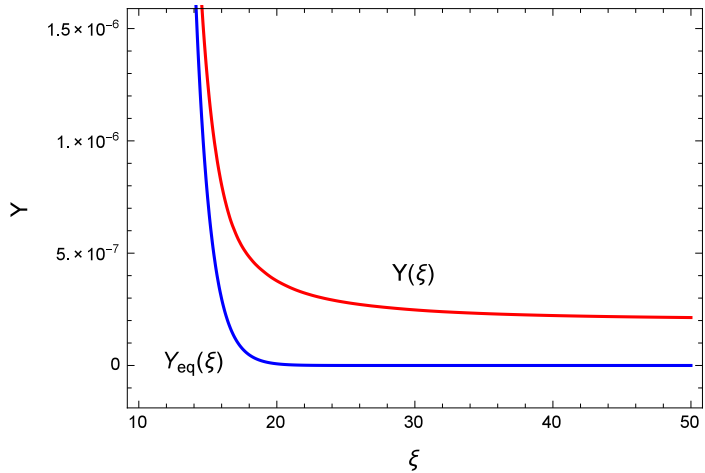}
\caption{\label{fig:2} The red curve shows the numerical solution $Y(\xi)$ of
Boltzmann equation~(\ref{Boltzmann-final}) for scalar millicharged particles and parameters
$\epsilon_5=5$, $\eta=1$ ($m_X=m_e$). The blue one is the equilibrium function, $Y_{eq}(\xi)$.}
\end{figure}

The plot of function $Y(\xi)$ for scalar $X$ and $m_X=m_e$ is presented in
Fig.~\ref{fig:2}. The picture for another mass and/or spin of $X$ would be similar:
when the temperature is high enough, the function  $Y(\xi)$ is close to the equilibrium one,
$Y_{eq}(\xi)$, but when the temperature drops below $\sim m_X/20$ the equilibrium is
upset and $Y(\xi)$ tends to the constant value, $Y_0$.

\begin{figure}[h]
\begin{tabular}{c c}
\includegraphics[width=0.49\textwidth]{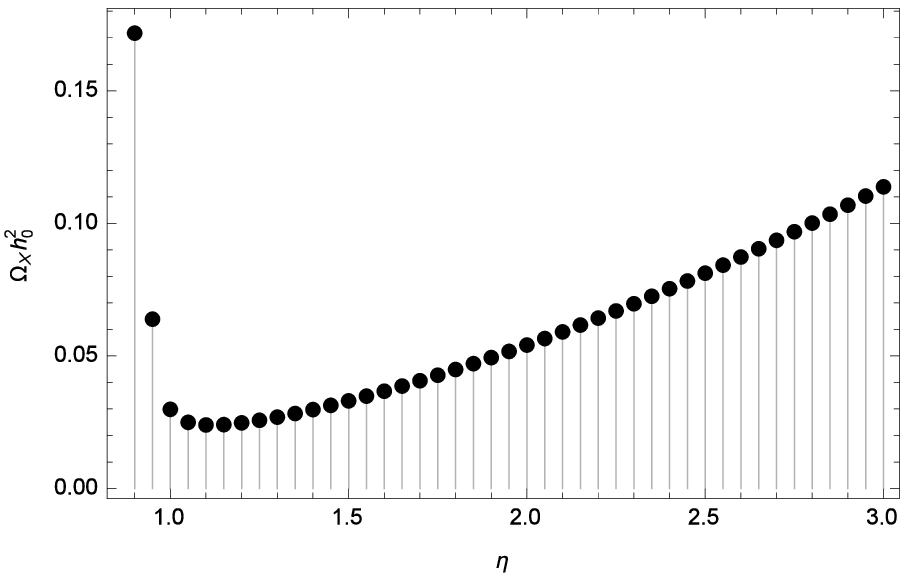} &
\includegraphics[width=0.49\textwidth]{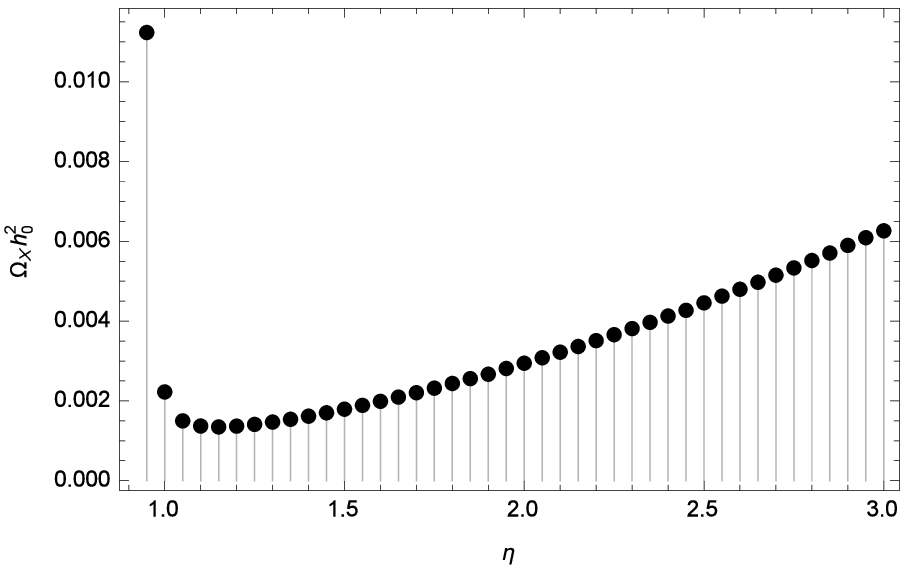}
\end{tabular}
\caption{\label{fig:3}
The relic abundance, $\Omega_X h_0^2$, of millicharged particles -- scalars (left plot) and
spin 1/2 fermions (right plot), for different values of the mass ratio, $\eta=m_X/m_e$, and for fixed $\epsilon_5=5$.}
\end{figure}

Calculating numerically $Y_0$ for different values of parameter $\eta$ and using
Eq.~(\ref{omega}) one can find $\Omega_X h_0^2$ for $X$-particles, both scalars and
fermions. The corresponding results are presented in Fig.~\ref{fig:3}.

The large values of $\Omega_X h_0^2$ for small $\eta$ have the simple explanation.
If annihilation $X\bar{X}\to e^+e^-$ proceeds "below threshold" ($m_X < m_e$), the
smaller is  $\eta$, the lower is the number of $X$-particles at the high energy tail of their
distribution which have enough energy to annihilate into heavier $e^+e^-$.
Therefore, $\Omega_X h_0^2$ grows exponentially fast for $\eta<1$, when $\eta$ decreases.

When $\eta>1$, all $X$-particles annihilate "above threshold". Therefore, $Y_0$
weakly depends on $\eta$ and remains almost constant, but $\Omega_X h_0^2$ in general
slightly increases with growth of $\eta$, mainly due to the factor $\eta$ in r.h.s. of
Eq.~(\ref{omega}). Therefore, it turns out that at $\eta\approx 1.1$ the cosmological fraction of MCPs,
$\Omega_X h_0^2$, takes the minimal value, when exponential decrease of function $\Omega_X h_0^2 (\eta)$
changes to its slow growth.
With rising $\eta$, the number density of MCPs behaves as $ n_X \sim 1/(\sigma_{ann} m_X) \sim m_X$
and hence their energy density rises as $m_X^2$, i.e. $\Omega_X h_0^2$ increases like $\eta^2$.

The values of $\Omega_X h_0^2$ for fermions are much smaller than those for scalars
at the same $\eta$ (see Fig.~\ref{fig:3}), because of behavior of the corresponding annihilation
cross-sections: since scalars annihilate in $P$-wave, their cross-section is suppressed near the
threshold by an extra power of velocity, $v \sim T/m_X$, while
this is not the case for fermions which annihilate in $S$-wave.

\section{Conclusion} \label{sect-conclud}

We have calculated the cosmological energy density, $\Omega_X h_0^2$, of millicharged particles (MCPs)
with masses $m_X\sim m_e$ and with the electric charge $e' = 5\cdot 10^{-5} e$
which is the maximal value of MCP charge allowed by SLAC experiment~\cite{SLAC}.
We have found that $\Omega_X h_0^2$ can vary in a wide range of values,
depending on the ratio $\eta=m_X/m_e$. For the subthreshold annihilation $X \bar X \to e^+ e^-$ ($m_X < m_e$)
it can be even as large as the observed energy density of dark matter $\sim 0.2$.
On the contrary, for $m_X \gtrsim m_e$ the cosmological energy density of MCPs
can be low enough, $\Omega_X h_0^2 \approx 0.02$ for scalar MCPs,
and $\Omega_X h_0^2 \approx 0.001$ for spin 1/2 fermions.

However, even the lowest value of $\Omega_X h_0^2$ obtained here either contradicts (for scalars)
or at least is in some tension (for spin 1/2 fermions)
with the most stringent CMB bound~\cite{dubovsky-2}, $\Omega_X h_0^2<0.001$ (95\% CL).
Therefore, it seems that in simple models millicharged particles (especially scalars) can not
contribute to the dark matter.

Nevertheless, in more complicated scenarios the possibility
that millicharged particles can constitute some part of dark matter still remains.
In particular, the CMB bound~\cite{dubovsky-2} can be considerably weakened
if the temperature of the relic MCPs is higher than the proton
temperature near the hydrogen recombination. This exotic possibility can be realized
if there exists a new long lived particle, which decays to $X\bar{X}$-pair prior to recombination
and heats them up. However, for an effective heating a new stronger interaction between MCPs
is necessary.

On the other hand, the SLAC bound~\cite{SLAC} on the value of millicharge may be relaxed
if new (anomalous) interactions of $X$-particles exist which could
strongly diminish their mean free path in matter, but more work is necessary to check compatibility of this
hypothesis with other particle physics data.

Therefore, it is still not completely forbidden that the MCPs with $m_X \sim m_e$
can be noticeable part of the cosmological dark matter. However, more detailed investigation
of the suggested and other exotics is surely needed to satisfy bounds from
the particle physics experiments.

\section*{Acknowledgements}
We acknowledge support of the Grant of President of Russian Federation
for the leading scientific Schools of Russian Federation,
NSh-9022-2016.2.

\end{document}